\documentclass{article}
\usepackage[T1]{fontenc} 
\usepackage[utf8]{inputenc} 
\usepackage{ismir,amsmath,cite,url}
\usepackage{graphicx}
\usepackage{color}
\usepackage{booktabs}
\usepackage{tabularx} 
\usepackage{lineno}
\usepackage{amssymb}
\usepackage{float}

\title{PerTok: Expressive Encoding and Modeling of Symbolic Musical Ideas and Variations}

\twoauthors
 {Julian Lenz} {Lemonaide Research \\ Barcelona, ES}
 {Anirudh Mani} {Lemonaide Research \\ Boston, US}

\def\authorname{Julian Lenz, and Anirudh Mani}

\usepackage[bookmarks=false,pdfauthor={\authorname},pdfsubject={\papersubject},hidelinks]{hyperref}

\begin{document}

\maketitle

\begin{abstract}

We introduce \textit{Cadenza}, a new multi-stage generative framework for predicting expressive variations of symbolic musical ideas as well as unconditional generations. To accomplish this we propose a novel MIDI encoding method, \textit{PerTok} (Performance Tokenizer) that captures minute expressive details whilst reducing sequence length up to 59\% and vocabulary size up to 95\% for polyphonic, monophonic and rhythmic tasks. The proposed framework comprises of two sequential stages: 1) \textit{Composer} and 2) \textit{Performer}. The Composer model is a transformer-based Variational Autoencoder (VAE), with Rotary Positional Embeddings (RoPE)\cite{ROPE} and an autoregressive decoder modified to more effectively integrate the latent codes of the input musical idea. The Performer model is a bidirectional transformer encoder that is separately trained to predict velocities and microtimings on MIDI sequences. Objective and human evaluations demonstrate Cadenza's versatile capability in 1) matching other unconditional state-of-the-art symbolic models in musical quality whilst sounding more expressive, and 2) composing new, expressive ideas that are both stylistically related to the input whilst providing novel ideas to the user. Our framework is designed, researched and implemented with the objective of ethically providing inspiration for musicians. 

\end{abstract}

\section{Introduction}\label{sec:introduction}

The creative endeavor in present-day music production is inherently complex and multifaceted. However, it can be broadly categorized into distinct phases that include 1) initiation, 2) evolution and development, and 3) completion of musical ideas into a finished musical outcome. Modern generative models have had a major impact in every creative domain, none the least in music creation. MIDI, and therefore Symbolic AI research approaches for single-track MIDI generation are especially applicable to the contemporary music producer. The motivation behind our investigation arises from a gap in this current landscape to facilitate the crucial middle phase of the creative process: absence of a comprehensive, adaptable modeling framework specifically engineered for generating expression variations from a given MIDI file input. Our proposed solution, \textit{Cadenza}, addresses this by focusing on the 'development' phase of music creation while unveiling a framework that is designed for flexibility and efficiency. Cadenza utilises a multi-stage generative process, the \textit{composer} and the \textit{performer}, to create novel ideas and variations while emulating the nuanced performance characteristics that can define a given musical style. We choose to call our framework 'Cadenza', inspired by the improvised musical passage played by soloists, creating new and exciting variations of the original motifs of the piece being performed.


\subsection{Encoding Symbolic Music} 
Alongside transformer-based architectures a number of methods have been proposed to encode, or \textit{tokenize} MIDI files into a discrete sequence of tokens. As transformers suffer from quadratic memory complexity in relation to sequence lengths \cite{AttentionIsAllYouNeed}, particular focus is placed on capturing relevant MIDI information whilst minimizing the total number of tokens. Popular tokenizers include REMI\cite{REMI}, TSD\cite{MidiTok} and Structured \cite{Structured}, amongst many others. However, these approaches suffer from a common drawback: they rely on singular tokens to denote the position of each note event on an evenly-spaced temporal grid. In comparison to MIDI files, which typically utilise a time resolution of 220 or 440 \textit{ticks-per-quarter} [note], these tokenizers are typically employed with just four intervals per quarter note. This has two negative effects: first, note-values outside this range are immediately \textit{quantized}, such as quarter-triplets, eighth-triplets, quintuplets, and thirty-second notes. In addition, any rhythmic performance attributes, expressed as subtle deviations from the fixed-grid, are immediately lost. As a result, the current state-of-the-art in MIDI tokenizers are unable to accurately capture the full range of rhythmic values and expressive performances.


\subsection{Expressive Modeling} 
In both digital and physical contexts, it is common to divide the act of \textit{composing} music and \textit{performing} it. The composition (or score) contains the raw musical idea, whereas the performance will typically embellish it with additional details, such as varying volumes (velocities in MIDI) and subtle timing deviations. Symbolic datasets can be categorized broadly as:

\begin{itemize}
    \item \textbf{Score}: The sequences contain quantized rhythmic values and minimal/no volume information. 
    \item \textbf{Time-Performance}: Dynamics and expressive timing are captured. The performer(s) play without a fixed tempo, resulting in a \textit{time}-based encoding (typically milliseconds), such as \cite{Maestro}. 
    \item \textbf{Beat-Performance}: Dynamics and expressive timing are captured. The notes are recorded in relation to a fixed tempo, with rhythmic expressivity occurring as deviations from the quantized beats. 
    
\end{itemize}

\noindent Although a substantial quantity of \textbf{score} datasets exist, there are significantly fewer in both \textbf{performance} categories. As a result, systems such as that proposed in \cite{ThisTimeWithFeeling}, wherein both composition and performance elements are jointly trained and predicted, are limited by this inequality. 

A number of recent models have been proposed to exclusively add performance elements, such as RenderingRNN\cite{RenderingRNN} and ScorePerformer\cite{ScorePerformer}. However, they rely on the prediction of \textit{tempo} tokens in alignment with the \textbf{Time-Performance} standard. This results in MIDI files that are still rhythmically \textit{quantized}, albeit with varying tempos. We posit that this approach is incompatible with the common production standards of many modern genres that instead rely on fixed tempos. 

The framework in Compose \& Embellish\cite{ComposeAndEmbellish} proposed a system of jointly training Lead-Sheet (score) and Performance models. With a modified REMI\cite{REMI} tokenization they demonstrated that the lead-sheet model could be pre-trained on a greater quantity of \textbf{score} data, and subsequently fine-tuned with the performer, on a smaller \textbf{performance} dataset. Similar to prior systems, they quantize rhythms to the nearest 16\textsuperscript{th} position, and instead predict \verb|[Tempo]| for expressive timing. Furthermore, due to the joint-conditioning training method, the compose model can lose certain capabilities from the fine-tuning process as it fits to the smaller performance dataset.


\subsection{Generating Variations}
A number of models have been proposed to solve variations-adjacent tasks. ThemeTransformer\cite{ThemeTransformer} utilises contrastive representation learning in a sequence-to-sequence framework to generate a melody and accompaniment that recurrently incorporates the original theme. The authors of Music FaderNets\cite{MusicFadernets} instead propose a \textit{style-transfer} task, wherein a number of high-level attributes can be applied to transform a polyphonic sequence. Our work most notably builds off of the model proposed in MuseMorphose\cite{MuseM}, which uses a novel \textit{in-attention} mechanism in a transformer-based VAE for generating attribute-controlled variations on symbolic data. However, these models are designed to predict long-form sequences (16+ bars) that do not contain any expressive information. \\


Overall, our key contributions to the field through this work are two-fold. Firstly, in section \ref{subsec: pertok} we introduce \textit{PerTok}, a novel MIDI encoding method that captures expressive details with required granularity while maintaining compact sequence lengths and manageable vocabulary sizes. 
PerTok is implemented with the MidiTok\cite{MidiTok} library, released open source and is compatible with any token-based sequential generation model. 
Secondly, the \textit{Cadenza} framework itself represents a significant leap forward as presented in section \ref{sec: model}, integrating the 'Composer' and 'Performer' models into a cohesive architecture that is researched and designed for the domain of AI-assisted music creation. Our framework offers a natural and intuitive way for musicians to create and modify music, which can be tailored to specific stylistic goals. Human evaluations showcase how Cadenza matches other state-of-the-art MIDI models in unconditional score generation quality, creates dynamic variations on input ideas, and sets a new standard for human-like expressive articulations.

\begin{figure*}[!ht]
  \centering
  \includegraphics[width=12.2cm]{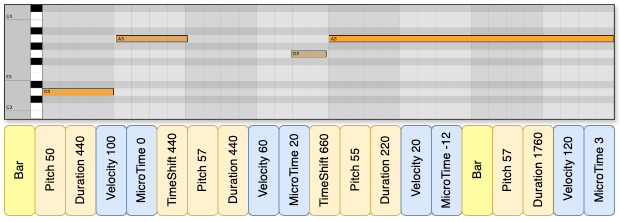}
  \caption{Example of PerTok tokenization on a 2-bar excerpt of a MIDI file. Composition tokens are highlighted in yellow, and performance tokens in blue.}
  \label{fig:pertok}
\end{figure*}

\section{Symbolic Data Encoding}\label{sec:symbolic data encoding}

We aim to encode MIDI data in a manner that is both 1) aligned with common audio production use-cases and 2) efficient in the context of transformer-based generative models. While a number of state-of-the-art models such as Anticipatory Music Transformer (AMT)\cite{AnticipatoryMT}, Figaro\cite{Figaro} and Multi-Track Music Machine (MMM)\cite{MMM} have focused on long-form, multi-track generation, we have observed from our experiences in designing products for contemporary music producers that they more commonly interact with short, single-track files. Furthermore, we observe that a number of tokenization methods such as \cite{ScorePerformer} rely on \textit{tempo} tokens to create a sense of expressive performance, whereas music producers often keep a singular, consistent tempo throughout their composition. Thus, our proposed encoding method is focused on single-track MIDI files in which the expressivity is calculated in relation to a fixed tempo. 

\subsection{Score \& Performance Encoding}\label{subsec:score and performance}

To address this, we create separate tokens to model the composition and performance timing elements. More specifically, the macro \textit{timeshift} tokens represent the quantized note locations within a score. Separately, the \textit{microshift} events denote a small adjustment from the quantized location, similar to the subtle timing deviations used by human performers. By separating these events, we are able to maintain reasonable balance between vocabulary size and total sequence length. Furthermore, the differentiation allows for models to be separately trained on the composition and performance tasks respectively. As there is a significant difference in the availability between quantized and performed symbolic datasets, this enables us to feed a far greater quantity of data into the composition-only model. 

Initial tests revealed that the common convention of quantizing the composition-level timeshift tokens to 16\textsuperscript{th} notes was leading to a number of musical issues. For example, when quarter- or eighth-note triplets were present in the input MIDI file, the quantization process was considering them to be 16\textsuperscript{th} notes with large degrees of \textit{microshift}. We addressed this by providing the ability to specify multiple, overlapping quantization grids, such as 16\textsuperscript{th}, quarter-triplets, and 8\textsuperscript{th}-triplets. Thus, the PerTok tokenizer is more adept at capturing the wide variety of rhythmic values that are commonly found in genres such as hip-hop, jazz and salsa (among many others).

\subsection{PerTok}\label{subsec: pertok}



\textbf{Score Tokens}: Similar to the MIDI-Like\cite{ThisTimeWithFeeling} and Structured\cite{Structured} encoding methods we represent \textit{macro} time changes between notes with \textbf{Timeshift} tokens. As MIDI time data is typically expressed as ticks-per-quarter, PerTok allows for multiple overlapping granularities to model a variety of rhythmic values. When encoding the MIDI data, PerTok matches each note's position to the closest possible timeshift value. \textbf{Pitch} is denoted as a MIDI pitch value between 0-127, with the capability to limit this range when musically appropriate. \textbf{Duration} tokens are used after each new note, to indicate the length of time before a MIDI note-off message is triggered. Notably, PerTok allows for the removal of duration tokens altogether, which helps further reduce sequence lengths when modeling rhythmic instruments.

\textbf{Performance Tokens}: With the addition of performance tokens, we aim to capture the musical subtleties that transform a written score into an expressive performance. \textbf{Velocity} tokens denote the strength of the note's attack, a property that is typically used in DAWs to augment timbre and volume characteristics. Although MIDI provides a range of 0 - 127 for velocity values, we allow for a bucketing approach to reduce a given velocity into one of \textit{n} possible values.  \textbf{Microshift} tokens provide a granular shift from the quantized rhythmic note value. PerTok is provided a maximum microshift value (e.g. 30 MIDI ticks) and a discrete number of possible microshift buckets. For example, \textit{Microshift 15} represents a placement of 15 ticks after the quantized note position, and \textit{Microshift 0} results in the initial quantized value.

In \textbf{Table \ref{table: pertok benchmark}} we provide a benchmark of our proposed PerTok encoding against a number of popular MIDI tokenizers. We sampled from 2,000 polyphonic 4- and 8-bar MIDI files that are used in modern audio production environments. For each tokenizer, we use 32 possible velocity buckets. To demonstrate the tradeoff between composition and performance, we initialize one version with 16th-note quantization (thus removing any performance characteristics), and a second (denoted with a -p) version with 440 timeshifts per quarter note. REMI and Structured methodologies use evenly spaced temporal grids to encode the location of each event, whereas PerTok uses a mixture of macro and micro timeshift tokens leading to a 95\% reduction in vocabulary size when encoding expressive rhythmic details. We additionally provide a visualization of our encoding method with a sample 2-bar melody in \textbf{Figure \ref{fig:pertok}}.


\begin{table}[]
\centering
\footnotesize
\begin{tabular}{@{}lrr@{}}
\toprule
\textbf{Tokenizer} & \multicolumn{1}{l}{\textbf{Vocab. Size}} & \multicolumn{1}{l}{\textbf{Seq. Length}} \\ \midrule
REMI              & 273  & 195 \\
REMI-p            & 5505 & 199 \\
Structured        & 289  & 216 \\
Structured-p      & 7265 & 216 \\
TSD               & 288  & 188 \\
TSD-p             & 7264 & 192 \\
\textbf{PerTok}            & 196  & 134 \\
\textbf{PerTok-p}          & 259  & 243 \\
\textbf{PerTok no-duration} & 164  & 80  \\ \bottomrule
\end{tabular}
\caption{Vocabulary sizes and average sequence lengths for popular tokenizers and our proposed PerTok encoding.}
\label{table: pertok benchmark}
\end{table}

\section{Model}\label{sec: model}

\begin{figure*}[]
  \centering
  \includegraphics[width=17.2cm]{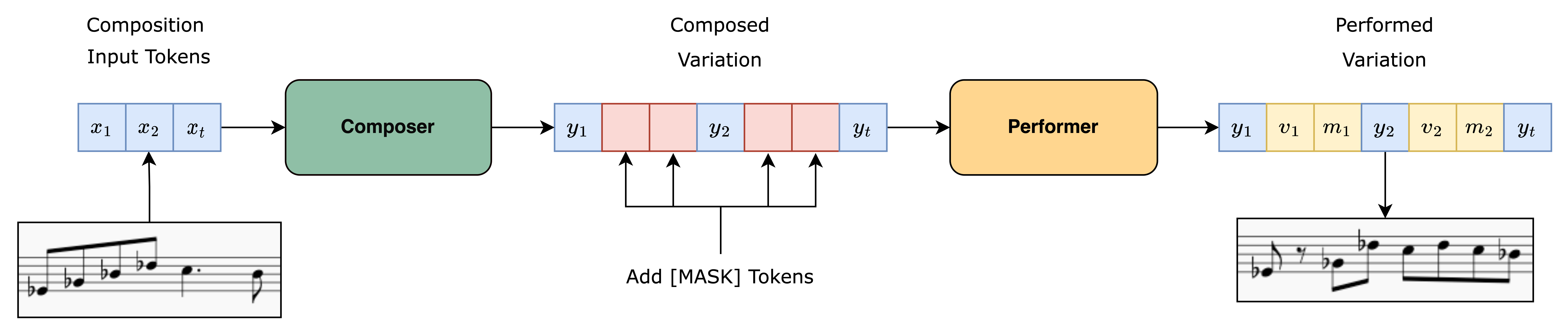} 
  \caption{An overview of the multi-stage \textit{Cadenza} architecture.}
  \label{fig:cadenza} 
\end{figure*}

With the objective of generating expressive, beat-structured variations on an input MIDI file, we present \textit{Cadenza}, a multi-stage VAE with transformer-based components. The framework is designed upon a principle that the \textit{composition} task requires a significant quantity of data and benefits from auto-regressive generation, whereas the patterns of expressive \textit{performance} tokens can be learned with smaller datasets and predicted in a bi-directional manner. 

\subsection{The Composer}
Given an input sequence of tokens \(\{x_1, x_2, ..., x_t\}\) in which \(x_t\) represents a single token \(x\) at index \(t\) the \textbf{composer} model is designed to auto-regressively predict \(\{y_1, y_2, ..., y_t\}\), an output sequence that is musically related yet distinct from the input. We utilise a sequence-to-sequence VAE architecture with in-attention conditioning similar to MuseMorphose\cite{MuseM}, enabling the model to learn a compressed, latent representation of short musical ideas within a regularized space. Within each attention mechanism, the query \(q\) and key \(k\) vectors at timesteps \(m, n\) are obtained with RoPE\cite{ROPE} for enhanced positional context:

\begin{equation}
    q_m^\top k_n = (R^d_{\Theta,m} W^q x_m)^\top (R^d_{\Theta,n} W^k x_n) 
\end{equation}

Wherein \(R^d_{\Theta,m}\) and \(R^d_{\Theta,n}\) are the rotary matrices for positions for embedding positional information, \(W^q\) and \(W^k\)  are learnable weight matrices transforming inputs \(x_m\) and \(x_n\) into the query and key vectors. For additional context we encourage readers to refer the original paper. As music is a deeply temporal phenomenon, the composer benefits from the increased token spatial modelling that is provided by the rotary embeddings. 

The encoder is designed to create a latent vector \(z\) of the input musical idea which serves as an anchor throughout the decoding process. In alignment with the original transformer\cite{AttentionIsAllYouNeed} we first project the input sequence with a learned embedding space, transforming it into \(X \in \mathbb{R}^{d \times t}\) where \(d\) is the hidden dimension size. This is then processed through several multi-head self-attention layers. Following a similar approach to \cite{MuseM} we extract the first timestep of the final attention layer output to obtain hidden vector \(h \in \mathbb{R}^{d}\), a contextual representation of the full input sequence. 

Following standard VAE methodology \cite{VAE}, the output vector is then processed through two learnable weight matrices \(W_{\mu} \in \mathbb{R}^{d \times d_z}\) and \(W_{\sigma} \in \mathbb{R}^{d \times d_z}\), wherein \(d_z\) denotes the size of the latent dimension. This process yields the mean \(\mu\) and  standard deviation \(\sigma\) vectors, encapsulating the latent space distribution parameters. Using the reparameterization trick, we sample \(\epsilon\) from the Gaussian distribution  to obtain \(z \in \mathbb{R}^{d_z}\) from our encoder:

\begin{equation}
    \begin{gathered}
        z = hW_{\mu} + hW_{\sigma} \odot \epsilon
    \end{gathered}
\end{equation}

The encoder's output distribution \(q(z|X)\) is aligned to a Gaussian prior \(\mathcal{N}(0, 1)\) by the traditional Kullback-Leibler (KL) divergence loss term:

\begin{equation}\label{equation:kl}
    D_{KL}(q(z|X) \| p(z)) = -\frac{1}{2} \sum_{k=1}^{d_z} \left( 1 + \log(\sigma_k^2) - \mu_k^2 - \sigma_k^2 \right)
\end{equation}

We further modify the equation utilising free bits as proposed by \cite{freebits}, allowing the encoder a degree of unpenalized space defined by \(\lambda\) to learn musical attributes without regularization.

\begin{equation}
    \mathcal{L}_\text{KL} = \sum_{k=1}^{d_z}\text{max}(\lambda,D_\text{KL}(q(z_k|X)||p(z_k)))
\end{equation}

The decoder is trained to autoregressively predict an output sequence of tokens whilst maintaining a recognizable connection with the input musical idea. Initially, the latent vector \(z\) is expanded to the decoder's hidden dimension \(d\) via a learnable matrix \(W_\text{pre} \in \mathbb{R}^{d_z \times d}\). We separately expand the input tokens \(v_1, v_2, ..., v_k\) with the same embedding layer used by the encoder.

In the context of the autoregressive VAEs it has been noted that \textit{posterior collapse} is a common issue \cite{PosteriorCollapse, MMD, BridgingMMD}, in which a sufficiently powerful decoder can simply ignore the encoder's regularized information, instead relying purely on the previous tokens. We utilise the methodology of Skip-VAE\cite{SkipVAE} and MuseMorphose\cite{MuseM}, integrating the proposed \textit{in-attention} mechanism. Prior to each attention layer, we sum expanded latent vector \(z_\text{pre}\) with every timestep of the previous hidden state, thus reinforcing its information throughout every stage of the decoding process. Therefor hidden state \(H \in \mathbb{R}^{t \times d}\) at layer \(i\) is calculated as:
\begin{equation}
    H_i = \text{SelfAttention}(H_{i-1} + x_\text{pre})
\end{equation}

The final hidden state is then passed through a feed-forward layer, which has weights tied to the embedding layers as first proposed in \cite{WeightTying}. The decoder minimizes the negative log likelihood (NLL) of the output sequence \(y_t\) when given prior tokens:

\begin{equation}
    \mathcal{L}_{\text{recon}} = \sum_{t=1}^{T}\text{log}p_\theta(y_t|y_{<t}, z)
\end{equation}

Thus, the composer is optimized with the NLL reconstruction loss as well as the \(\beta\)-scaled regularization KL loss:

\begin{equation}
    \mathcal{L_\text{composer}} = \mathcal{L}_\text{recon} 
    + \beta\mathcal{L}_\text{KL}
\end{equation}

\subsection{The Performer}\label{sec:Performer Model}

The Performer is separately tasked with computing tokens for \textbf{velocity} and \textbf{microshift} time values. Thus, it is able to transform a quantized MIDI score into one with expressive characteristics, rendering it more suitable for a variety of music production tasks. Whereas the composition task generally benefits from an autoregressive setup wherein each token is predicted sequentially, it has been demonstrated in \cite{Tap2Drum, VAEDER} that performance attributes can be predicted in a bi-directional manner. 

We utilise a framework comparable to the masked token prediction task of BERT\cite{BERT}. During training and inference, we replace the input performance tokens denoting velocity and microtiming with a single \verb|[MASK]| token. The tokens related to pitch, timeshift, and duration are left unmodified. The model is a standard transformer encoder as per \cite{AttentionIsAllYouNeed}, with sinusoidal positional embeddings, layer normalization and a final feedforward layer that has weights tied to the initial embedding layer. 

During training the model is tasked to replace each \verb|[MASK]| token with an appropriate velocity and microshift value, with cross-entropy loss used exclusively on the masked tokens. We perform the masking operation on 100\% of performance tokens. At inference time, we manually mix tokens between the original source and model predictions, thus ensuring the original pitch, timeshift and duration values are maintained.

\section{Experimental Setup}\label{sec:experiments}


\subsection{Composer Ablations}

By training several composer models, we aim to understand the relationship between various degrees of KL regularization and the decoded sequence's similarity to the input. Each model is trained on 4-bar segments of the full Lakh-MIDI dataset\cite{LakhMIDI}. The models all have 12 layers and 8 heads in both the encoder and decoder, with a latent dimensionality \(d_z\) of 128 and hidden dimension \(d\) of 512. The \textit{Full-KL} model was trained with a KL regularizer \(\beta=1.0\) and free bit \(\lambda=0.15\). The \textit{Balanced-KL} model was trained with \(\beta=0.3\) and \(\lambda=0.25\). In both instances the KL regularization was applied with cosine cyclical annealing\cite{CyclicalAnnealing} every 10,000 steps. We initially keep \(\beta=0.0\) for the first 25,000 steps, and then linearly raise it to the maximum value over the proceeding 25,000 steps. Finally, the \textit{No-KL} model had a \(\beta=0.0\) (no regularization), thus allowing the encoder to exclusively optimize against reconstruction quality. 

\textbf{Objective Evaluations} : We generate a single variation for 500 files from the test set for each model, utilising greedy decoding to remove any sampling logic from the evaluation framework. For each sample, we calculate the similarity in \textbf{pitch distribution}, \textbf{onset locations} and \textbf{durations}:

\begin{equation}
    similarity(x^a, x^b) = 100 \frac{\langle x^a, x^b \rangle} 
    {||x^a|| ||x^b||}
\end{equation}

\noindent wherein \(x \in \mathbb{Z}^t \) is a discrete vector of \(t\) attributes; in the case of \textbf{pitch} we set \(t = 128\) to capture the full MIDI note range, and for both \textbf{onset location} and \textbf{duration} \(t = 64\), representing the nearest 16\textsuperscript{th} value in a 4-bar pattern. Finally, we report \textbf{Absolute Similarity}, the percentage of notes that have \textit{identical} characteristics (pitch, onset, duration) between both sequences.

\subsection{Performer Fidelity}

Two performer models are trained with separate datasets to measure their capacity to model the unique expressive characteristics of a given training set. Both models are trained with 12 layers and heads, a hidden dimensionality of 768, and a dropout of 10\%. One model is trained on the classical MusicNet dataset \cite{MusicNet}, and the other (referred to as \textit{HipHop}) is trained on a proprietary hip-hop dataset. In both cases, we train on approximately 10,000 4-bar excerpts. Each dataset contains polyphonic data with differing expressive patterns of velocities and microtiming. 

We randomly extract 2,000 polyphonic 4-bar patterns from the Lakh-MIDI dataset \cite{LakhMIDI} and generate expressive tokens from both models. We subsequently measure the velocity and microtiming distributions of both the generations as well as the two original training datasets. \textbf{Velocity} distribution is represented as \(v \in \mathbb{Z}^{128}\), a vector representing the number of occurences of each velocity value. For each note, \textbf{microtiming} is calculated as a percentage deviation from the nearest 16\textsuperscript{th} note, with +/-50\% denoting the halfway point to/from the adjacent 16\textsuperscript{th} time index. The total distribution of \textbf{microtiming} deviations in a given sequence is thus represented with vector \(mt \in \mathbb{Z}^{100}\). 

For both velocity and microtiming, we compare the distributions of both model's predictions against the MusicNet and HipHop datasets. In \textbf{Table \ref{tab:performer_fidelity}} we report the KL divergence, as well as the absolute difference in the mean and standard deviation for these distributions. In each metric, a lower value indicates higher degree of similarity between the model's predictions and the original dataset's expressive characteristics.

\subsection{User Study}

We conduct a thorough user study to achieve a qualitative understanding of our model’s performance, comparing to different external baselines and versions due to hyperparameter settings. All the audio samples that users heard were 4 bar MIDI files voiced through the same Piano VST. In Part A of the user study, 25 human evaluators listened to 5 seed melodies and 4 alternative variations for each of them, coming from the \textit{No-KL}, \textit{Balanced-KL} and \textit{Full-KL} versions of our proposed model, and additionally a \textit{Placebo} melody which was randomly selected to be in the same key and scale as the input but had no relation to it. This was done to confidently ground the musical understanding of our human evaluators. Overall, 44\% of our evaluators identified themselves as "Novice : I have little to no experience making music", 32\% as "Amateur : I love making music for fun",  and 24\% as "Professional : I regularly make music in a professional capacity".  

In Part B, the same 25 human evaluators also rated 3 unconditional generations from Cadenza, Anticipatory Music Transformer (AMT)\cite{AnticipatoryMT}\footnote{Specifically, the \href{https://huggingface.co/stanford-crfm/music-medium-800k}{music-medium-800k} checkpoint} and Figaro\cite{Figaro} models, which broadly represent the state-of-the-art in symbolic polyphonic generation. We randomly sampled 3 generations from publicly-available checkpoints of each model, all of which were trained on identical versions of the Lakh MIDI dataset\cite{LakhMIDI}, and present the results in \textbf{Table \ref{tab:huma_eval_model_scores}}. As Cadenza's composer requires a latent vector, we randomly sample from a Gaussian distribution for unconditional generations. For each sample, the evaluator was asked to rate between how musically appealing it sounded to them with 1 being the lowest and 4 being the highest score. In addition, they were also asked to select, in a binary choice, whether they thought the performance was generated by a human or computer.

\begin{table}[]
\centering
\resizebox{\columnwidth}{!}{%
\begin{tabular}{@{}lrrrrr@{}}
\toprule
\textit{Model} & \multicolumn{1}{l}{Pitch Sim.(\%)} & \multicolumn{1}{l}{Onset Sim.(\%)} & \multicolumn{1}{l}{Duration Sim.(\%)} & \multicolumn{1}{l}{Absolute Sim.(\%)} & \multicolumn{1}{l}{Human Eval(1-5)} \\ \midrule
Full-KL        & 71.01                          & 77.99                          & 88.65                             & 9.44                              & 1.94 \\
No-KL          & 95.64                          & 83.04                          & 98.40                             & 20.05                             & 2.84 \\
Balanced-KL    & 92.06                          & 80.80                          & 96.95                             & 16.46                             & 2.71 \\
\textit{Placebo} & -                              & -                              & -                                 & -                                 & 1.22 \\ \bottomrule
\end{tabular}%
}
\caption{Objective and human-evaluation results from the ablation studies. Higher values indicate more similarity to the input's musical characteristics.}
\label{tab:ablation_objective_evals}
\end{table}

\begin{table}[ht]
\centering
\resizebox{\columnwidth}{!}{%
\begin{tabular}{lcc}
\toprule
Model (Params) & Musical Appeal Score $\uparrow$ (1-4) & `Human-like' Score $\uparrow$ (1-2) \\
\midrule
AMT (360M) & \textbf{3.33} & 1.57 (57.3\%) \\
Figaro (87M)      & 2.73 & 1.57 (57.3\%) \\
\textbf{Cadenza} (142M)   & 2.91 & \textbf{1.72 (72.0\%)} \\
\bottomrule
\end{tabular}
}
\caption{Human Evaluation Results for Model Quality. Percentages represent the fraction of modeling outputs that were selected by human evaluators when asked if it could have been created by a human.}
\label{tab:huma_eval_model_scores}
\end{table}




\section{Results and Discussion}

We discuss our experimental results to answer three high level questions - 1) provided an input MIDI sequence, how \textit{musically related} are the Cadenza variations; 2) can the performer model tangibly improve expressivity; and 3) how appealing are the novel generations. We analyze our proposed scientific approach through quantitative and qualitative measures. 

Both human and objective evaluations demonstrate in \textbf{Table \ref{tab:ablation_objective_evals}} that training the composer with varying degrees of KL regularization has noticeable impacts on the balance between between recall and variety. Provided a musical idea as input, the \textit{No-KL} model will produce outputs nearly identical to the input melody. Alternatively, the \textit{Balanced-KL}  model will produce outputs that are related, yet altered enough to provide new sources of inspiration. In many cases, the \textit{Full-KL} model will produce entirely unrelated outputs, as a result of the encoder's heavy focus on regularizing the latent vectors. Since a score of 4.0 for a generation would be considered identical we can infer that the ideas generated by \textit{No-KL} and \textit{Balanced-KL} are roughly 70\% related to the input. This aligns with the quantitative results, which consistently show a negative correlation between the KL regularization and input/generation similarity metrics. As such, our framework is demonstrated to consistently generate variations that are perceptually relevant to, yet distinct from, the input musical idea. 

In \textbf{Table \ref{tab:performer_fidelity}}, we report results from the Performer Fidelity quantitative study. In both MusicNet and HipHop models, the distributions of predicted velocities and microtimings are consistently closer to that of their respective training datasets. We can therefore infer that the performer model, in conjunction with our newly proposed PerTok tokenizer, is capable of accurately learning the patterns of expressive characteristics from a comparatively small dataset. 

In \textbf{Table \ref{tab:huma_eval_model_scores}} we compare Cadenza to AMT\cite{AnticipatoryMT} and Figaro\cite{Figaro} on the task of novel generations. Our model, although comparable to its competitors on musical appeal, comes in second to the AMT. However, Cadenza comprehensively outperforms other models by 14.7\% in the \textit{human-like} expressivity ratings. We note that our framework is highly adaptable, in that both composer and performer models could be replaced with any type of sequential network. Theoretically, one could further improve the unconditional generation quality by replacing our VAE-based composer model with a decoder-only model, similar to that of AMT. 


\begin{table}[]
\resizebox{\columnwidth}{!}{%
\begin{tabular}{lrrrrrr}
\hline
Model (Metric)           & \multicolumn{2}{c}{KL} & \multicolumn{2}{c}{Mean $\Delta$} & \multicolumn{2}{c}{Std Dev $\Delta$} \\ \hline
\textit{} &
  \multicolumn{1}{c}{Train} &
  \multicolumn{1}{c}{Opposite} &
  \multicolumn{1}{c}{Train} &
  \multicolumn{1}{c}{Opposite} &
  \multicolumn{1}{c}{Train} &
  \multicolumn{1}{c}{Opposite} \\
HipHop (Velocity)       & \textbf{1.68}  & 3.63  & \textbf{1.81}      & 16.83      & \textbf{1.74}        & 9.04        \\
HipHop (Microtiming)     & \textbf{0.66}  & 2.64  & 0.05               & 0.05       & \textbf{0.00}        & 0.14        \\
MusicNet (Velocity)    & \textbf{3.17}  & 11.57 & \textbf{1.95}      & 13.06      & \textbf{1.20}        & 12.00       \\
MusicNet (Microtiming)  & \textbf{0.07}  & 3.17  & \textbf{0.02}      & 0.13       & \textbf{0.01}        & 0.15        \\ \hline
\end{tabular}%
}
\caption{Objective results on the Performer fidelity evaluations.}
\label{tab:performer_fidelity}
\end{table}

\section{Conclusion}

We introduced a multi-stage generative framework which allows for both variation and novel generation tasks, maintaining a competitive quality of composition and setting a new state-of-the-art of expressive characteristics. Our proposed tokenizer is used to create expressive symbolic sequences while effectively reducing vocabulary size and sequence length. We invite readers to our model page \footnote{Access code and demonstrations on our model page \href{https://lemo123.notion.site/ISMIR-Cadenza-Demo-Page-7028ad6ac0ed41ac814b44928261cb68?pvs=4}{here}. PerTok tokenizer is available as part of MidiTok library \href{https://miditok.readthedocs.io/en/latest/tokenizations.html}{here}.} where we showcase its fidelity in generating outputs for polyphonic, monophonic, bass and drum instruments.

In particular, our performer model in conjunction with the new tokenization method led to a quantifiable increase in listener's perceptions of the expressivity in the generations. These results were achieved with relatively tiny datasets, paving the way for further collaborations with artistic communities. Future research directions include exploring \textit{controllability}, as well as further improvements in the domain of novel generation.

\section{Ethics Statement}



As generative AI technology advances rapidly, it is crucial to address the implications of these developments in the generative domain. Concerns such as perpetuating cultural biases, undermining artists' financial opportunities, and using data without proper consent require urgent attention and dialogue within research communities. When developing new models, we must carefully consider both their intended applications and potential impacts.

Our research involves deep collaboration with artists to understand their motivations and needs, ensuring our efforts benefit the creative communities we serve. For instance, our new framework, designed for the MIDI symbolic domain, focuses on enhancing artists' tools with features that inspire creativity rather than replacing the artists. We also deliberately chose to work with smaller models, which helps minimize data requirements. This strategy promotes fair data agreements and increases the chances of fairly compensating musicians, thus fostering sustainability in creative industries and prioritizing ethical responsibility, especially in creative domains.


\newpage
\bibliography{ISMIRtemplate}

\end{document}